\documentclass[letterpaper]{article}

\usepackage{natbib,alifeconf}  
\usepackage{amsmath}
\usepackage{url,hyperref,cleveref}
\usepackage{booktabs}
\usepackage{mathrsfs}
\usepackage{txfonts}
\usepackage{multirow}
\usepackage{graphicx}
\usepackage{tabularx}
\usepackage{subfig}
\usepackage{float}
\usepackage{afterpage}

\title{
Reflexive Composition of Elementary State Machines,\\
with an Application to the Reversal of Cellular Automata Rule 90}

\author{
    Chris Salzberg$^{1,*}$
    and
    Hiroki Sayama$^{2}$ \\
    \mbox{}\\
    $^1$Shopify \\
    $^2$Binghamton University, State University of New York
}

\begin{document}

\maketitle

\begin{abstract}
We explore the dynamics of a one-dimensional lattice of state machines on two states and two symbols sequentially updated via a process of ``reflexive composition.'' The space of 256 machines exhibits a variety of behavior, including substitution, reversible ``billiard ball'' dynamics, and fractal nesting. We show that one machine generates the Sierpi\'nski Triangle and, for a subset of boundary conditions, is isomorphic to cellular automata Rule 90 in Wolfram's naming scheme. More surprisingly, two other machines follow trajectories that map to Rule 90 in reverse. Whereas previous techniques have been developed to uncover preimages of Rule 90, this is the first study to produce such inverse dynamics naturally from the formalism itself. We argue that the system's symmetric treatment of \textit{state} and \textit{message} underlies its expressive power.
\end{abstract}


\let\thefootnote\relax\footnotetext{ $^{*}$ Corresponding author: \texttt{me@chrissalzberg.com}}

\newcommand{\thefootnote}{\arabic{footnote}}

\section{Introduction}

The question of how open-ended complexity can emerge from simple rules is foundational to the fields of both Complex Systems and Artificial Life. Models employed to reproduce such emergence commonly take the form of a discrete set of rules applied to a finite lattice of states. Cellular Automata (CA), the most popular of these models, have been shown to exhibit a surprisingly rich variety of evolutionary and lifelike behavior \citep{SayamaNehaniv2025}. Elementary Cellular Automata (ECA), two-state CA that evolve on a one-dimensional lattice, are arguably the most basic and minimal example of the emergence of complexity from simple, discrete rules \citep{Wolfram2002}.

Implicit in these popular models is a divide between, on the one hand, the \textit{states} of a system and, on the other hand, the \textit{rules} that apply to those states. The earliest abstract model of computation, the Turing Machine, embodies this distinction in its division of the world into \textit{machine} and \textit{tape} \citep{Turing1937}; Church's \(\lambda\)-calculus contains a similar divide between the \textit{function} and its \textit{argument} \citep{Church1936}. There is a conundrum, however, in the fact that we find no physical equivalent in nature to the rules and states of these artificial systems. Despite intriguing parallels, there are no cleanly distinguishable ``machine molecules'' or ``tape molecules'' in the complex mechanisms of DNA/RNA transcription, for example. Indeed, the assumption that such distinct categories \textit{must exist} limits the capacity of these systems to model their emergence from within the given frame of reference.

The notion of \textit{emergence of function}, unrepresented at the core of these rule-based systems, is nevertheless of central importance in drawing analogies between computational models and the physical world. The lack of a model to express such emergence inspired the ``algorithmic chemistry'' model of Fontana and Buss, which employs LISP expressions that interact algorithmically in a fixed ensemble \citep{Fontana1991,Fontana1994}. The notion has a powerful interpretation in molecular biology, where it serves as an alternative to dogmatic theories of information ``carriers'' and ``processors'' \citep{Wills1989}. Origin-of-life problems, in which the distinction between carriers and processors is not yet well defined, are an area where the study of such emergence in a computational model could provide valuable insight. The problem space, however, has eluded widespread attention and remains largely underdeveloped.

In this paper we explore the dynamics of a discrete system defined by a one-dimensional lattice of states and a finite-state machine applied to update them. Unlike cellular automata, the update process, referred to as ``reflexive composition,'' is symmetric in its treatment of \textit{state} and \textit{message}. This symmetry results in the surprising finding that certain elementary cellular automata, in particular Rule 90 in Wolfram's naming scheme, are represented in the system alongside their inverse (time reversed process). This inverse process is not expressible as cellular automata but arises naturally from the formulation presented. Furthermore, we show that for this rule, \textit{time reversal} is equivalent to the inversion of \textit{states} and \textit{messages} in the system's FSM. Interpretations of this finding are presented in the Conclusions section.

\section{Formulation}

The formulation of the ``composite state machine'' was introduced in earlier studies on a graph-based artificial chemistry \citep{Salzberg2006,Salzberg2006_MachineTape,Salzberg2007}. The focus in those studies was on state-space complexity; here our focus will be on exploring \textit{trajectories} in the state space, an aspect overlooked in previous work. In this section, we review the formulation, introduce a naming scheme for minimal state machines, and present a new visualization of trajectories that makes it easier to see the similarities with a subset of ECA.

\subsection{The finite state machine}

The basis for the formulation is the traditional finite state machine (FSM) from computer science theory \citep{Minsky1967}. A FSM is described in terms of a state-transition graph \(G = (Q, \mathscr{A}, \delta, \varphi)\), where \(Q\) is a set of states, \(\mathscr{A}\) a finite alphabet, \(\delta : \mathscr{A} \times Q \rightarrow Q\) a transition function, and \(\varphi : \mathscr{A} \times Q \rightarrow \mathscr{A}\) an output function. We refer to the transition and output functions for an input message \(i\in\mathscr{A}\) from a state \(q\in Q\) in the forms \(\delta _q (i)\equiv\delta(i,q)\) and \(\varphi_q(i)\equiv\varphi(i,q)\), respectively. The FSM constitutes a minimal abstraction of a physical machine: a collection of states, transitions between states, and elementary input/output mappings. Given an input \(i\in\mathscr{A}\), a FSM in state \(q\in Q\) follows the transition to state \(\delta _q(i)\) and returns an output \(\varphi _q(i)\).

\begin{figure}
    \centering
    \includegraphics[width=0.47\textwidth]{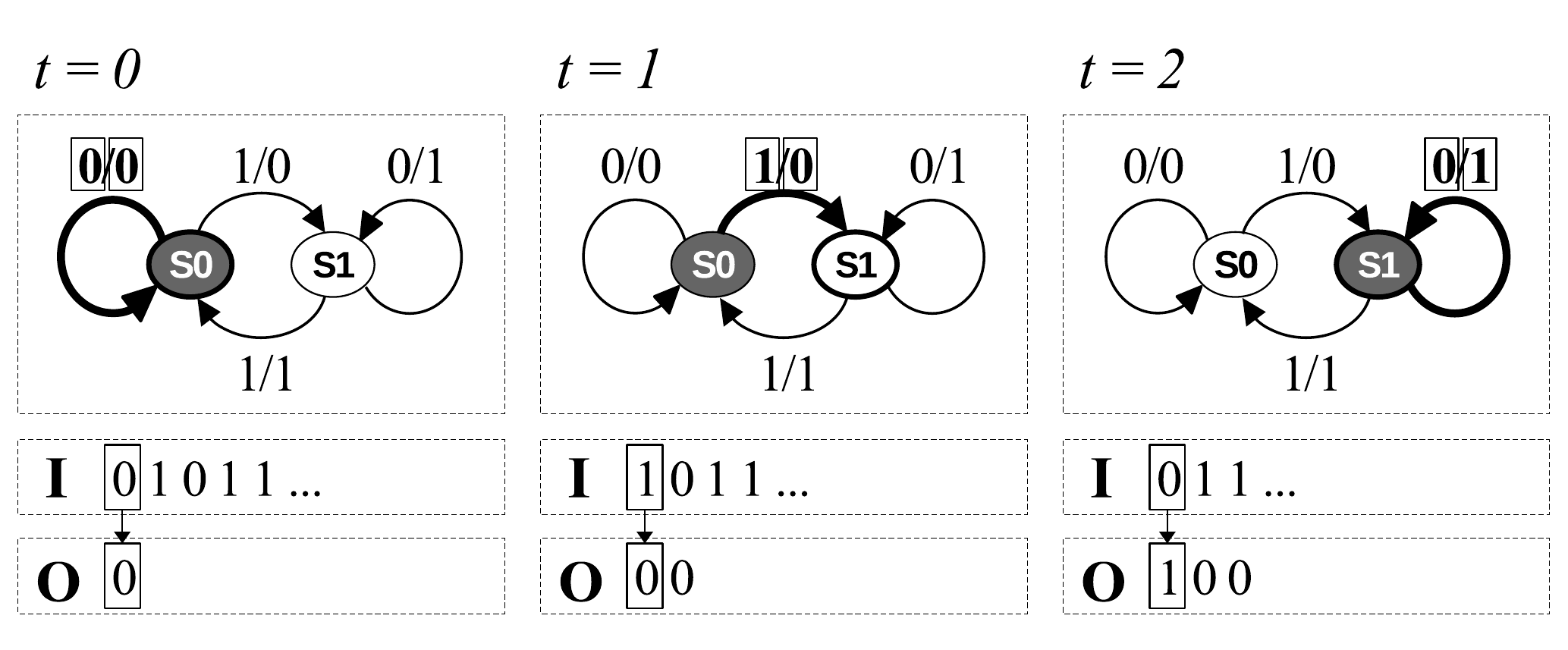}
    \caption{Three steps of FSM with \(Q = \{\textit{S0}, \textit{S1}\}\) and \(\mathscr{A} = \{0, 1\}\). State-transition function \(\delta\) and output function \(\varphi\) are defined in \Cref{M45}(a). Dark nodes and bold edges indicate the current state and transition, respectively, in each frame, with input and output streams shown below.}
    \label{fsm}
\end{figure}

Here we consider the evolution of an FSM over time, with \(i(t) \in\mathscr{A}\) representing an input stream, \(q(t)\in Q\) a dynamic state, and \(o(t) \in\mathscr{A}\) an output stream. The next state is then defined as \(q(t+1) \coloneqq\delta _q(i(t))\), and the output as \(o(t) \coloneqq\varphi _q(i(t))\). The evolution of a FSM is shown in \Cref{fsm} for the first three inputs \((\textit{0}, \textit{1}, \textit{0})\) of an input sequence.

\subsection{Naming scheme for elementary state machines}

\begin{figure}[ht]
\centering
\setlength{\tabcolsep}{0.5em}
\begin{tabular}{|c||c|c|c|c|c|c|c|c|}
\multicolumn{8}{c}{\includegraphics[width=0.3\textwidth]{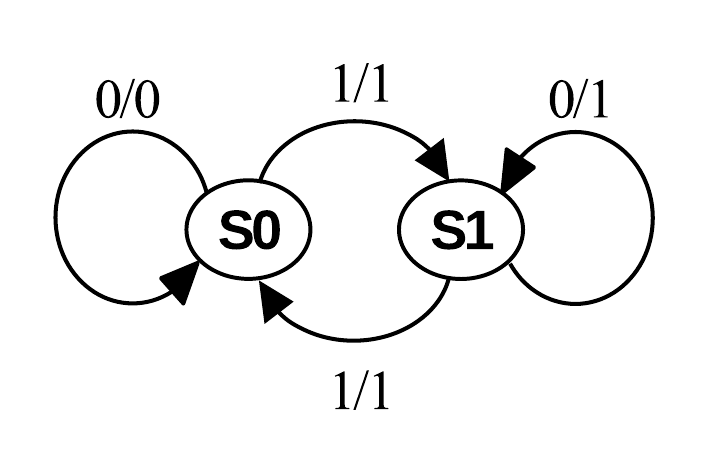}} \\
\hline
State &
\multicolumn{4}{c|}{\textit{S0}} & \multicolumn{4}{c|}{\textit{S1}} \\
\hline
Input &
\multicolumn{2}{c|}{\(\textit{0}\)} & \multicolumn{2}{c|}{\(\textit{1}\)} & \multicolumn{2}{c|}{\(\textit{0}\)} & \multicolumn{2}{c|}{\(\textit{1}\)} \\
\hline
{\small Destination / Output } &
{\small \textit{S0} } & {\small \(\textit{0}\) } & {\small \textit{S1} } & {\small \(\textit{1}\) } & {\small \textit{S1}} & {\small \(\textit{1}\)} & {\small \textit{S0}} & {\small \(\textit{1}\)} \\
\hline
\hline
Encoding &
{\small \texttt{0}} & {\small \texttt{0}} & {\small \texttt{1}} & {\small \texttt{1}} & {\small \texttt{1}} & {\small \texttt{1}} & {\small \texttt{0}} & {\small \texttt{1}} \\
\hline
\end{tabular}
\caption{FSM with identifier \textit{M61}. States \(\textit{S0}\) and \(\textit{S1}\) and message values \(\textit{0}\) and \(\textit{1}\) are each mapped to the bits \texttt{0} and \texttt{1}, respectively, resulting in the bit string (\(\texttt{00111101})_2 = 61\).}
\label{M61}
\end{figure}

To simplify the presentation of results, we introduce a naming scheme for the elementary state machines to be explored in later sections. Whereas previous work considered more complex state machines, our focus here will be on the minimal case of two states \(Q = (\textit{S0},\textit{S1})\) and two symbols \(\mathscr{A} = (\textit{0}, \textit{1})\). We can enumerate such machines by considering that for any particular machine, we must define state transition functions (\(\delta_{\textit{S0}}\) and \(\delta_{\textit{S1}}\)) and output functions (\(\varphi_{\textit{S0}}\) and \(\varphi_{\textit{S1}}\)) for two states, each taking and producing one of two possible values (\((\textit{0}, \textit{1})\) or \((\textit{S0},\textit{S1})\)). Thus each function requires two bits to represent, and we have four such functions to define, for a total of 8 bits of information and 256 possible machines. This is shown in \Cref{fsmdef}, with an example FSM, \textit{M61}, shown in \Cref{M61} along with its encoding.

Although there are a total of 256 possible elementary state machines, many are equivalent computationally to others in the same set. The choice of the message values \(\textit{0}\) and \(\textit{1}\) is arbitrary, so each machine has a ``complementary machine'' with these values reversed. Likewise, the states of a machine can be reversed without changing the dynamics of the system, yielding an equivalent ``mirror machine''. Finally, both message values and states can be swapped to yield a ``complementary mirror machine.'' Thus any FSM is computationally equivalent to as many as three other machines.

Of the set of all 256 elementary FSMs, only 76 are unique under these transformations. This is similar to the equivalence among ECA rules, of which 88 are computationally unique. In \Cref{canonicals}, we list the interesting FSMs of focus in this study and, where applicable, their equivalent CA rules.

\begin{table}[ht]
\centering
\begin{tabular}{|c||c|c|c|c|c|c|c|c|}
\hline
State &
\multicolumn{4}{c|}{\textit{S0}} & \multicolumn{4}{c|}{\textit{S1}} \\
\hline
Input &
\multicolumn{2}{c|}{\(\textit{0}\)} & \multicolumn{2}{c|}{\(\textit{1}\)} & \multicolumn{2}{c|}{\(\textit{0}\)} & \multicolumn{2}{c|}{\(\textit{1}\)} \\
\hline
Function &
\(\delta\) & \(\varphi\) & \(\delta\) & \(\varphi\) & \(\delta\) & \(\varphi\) & \(\delta\) & \(\varphi\) \\
\hline
Encoding Bit &
{\small 7} & {\small 6} & {\small 5} & {\small 4} & {\small 3} & {\small 2} & {\small 1} & {\small 0} \\
\hline
\end{tabular}
\caption{Naming scheme for two-state, two-input FSM. The results of the functions \(\delta\) and \(\varphi\) for each pair of arguments are concatenated into a binary string and converted to decimal, similar to the naming of ECA rules.}
\label{fsmdef}
\end{table}

\subsection{Reflexive composition}

With the FSM and its naming scheme defined, we now consider a process by which to evolve a set of states on a lattice. Consider first a pair of states \(q_s(t)\) (sender) and \(q_r(t)\) (receiver) in \(Q\); this will later be generalized to a lattice of arbitrary many states. In contrast to previous work, which considered the more general case of distinct sender and receiver state machines, here we assume the lattice is uniform, which will simplify our presentation. Sender and receiver states are thus both updated using the same FSM.

At each step, we calculate the output for the pair of states by composing the output functions at \(q_s\) and \(q_r\), \(\varphi_s\) and \(\varphi_r\), to produce the lattice (row) output \(o(t)\) for an input \(i(t)\):

\begin{equation}
\label{output}
o(t) = \varphi_r(\varphi_s(i(t)))
\end{equation}

\noindent Thus the output of the state pair is calculated simply by ``piping'' the output of the sender to the input of the receiver.

The state transition update employs a similar composition, incorporating the transition functions \(\delta_s\) and \(\delta_r\) at \(q_s\) and \(q_r\), and the output function \(\varphi_s\), to produce:

\begin{align}
\label{transition}
\begin{split}
q_r(t+1) &= \delta_s(i(t)), \\
q_s(t+1) &= \delta_r(\varphi_s(i(t))).\\
\end{split}
\end{align}

\noindent These equations advance each state in the lattice (here, \(q_s\) and \(q_r\)) along transitions corresponding to the input \(i(t)\) and its composition \(\varphi_s(i(t))\), passed as arguments to \(\delta_s\) and \(\delta_r\), respectively.

On the left side of \cref{transition}, we assign the results of these transitions to the next step states \(q_s(t+1)\) and \(q_r(t+1)\), but in doing so alternate the identity of sender and receiver by passing the result of \(\delta_s\) to \(q_r\) and \(\delta_r\) to \(q_s\). The alternation of sender and receiver in the update process, detailed extensively in \cite{Salzberg2006_MachineTape,Salzberg2007}, is essential to ensuring that no element of the system acts exclusively as information processor or information carrier. We refer to this symmetry between sender and receiver as ``reflexivity''.

\begin{figure}
    \centering
    \includegraphics[width=0.44\textwidth]{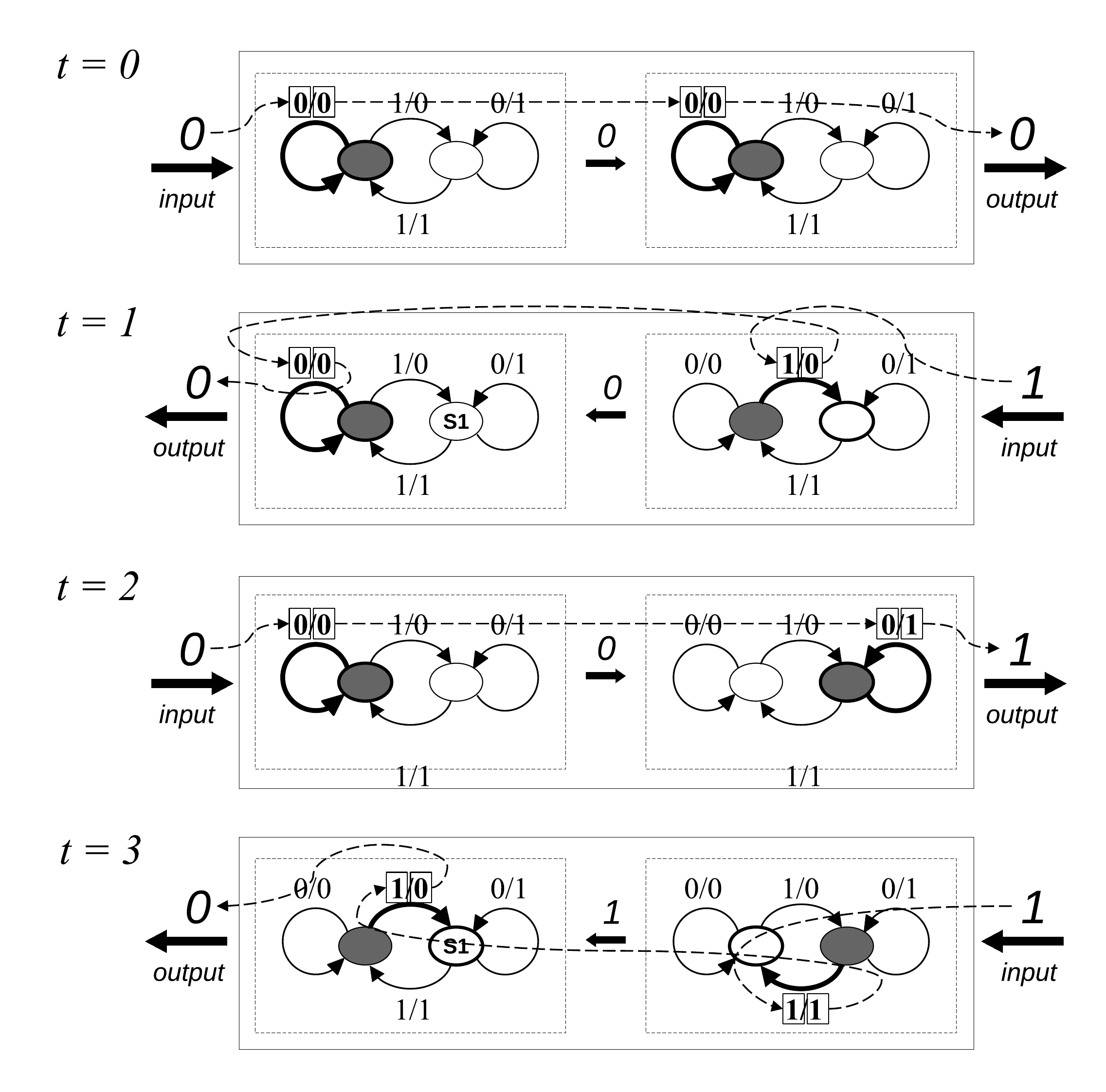}
    \caption{Four steps (vertical) of state composition (horizontal) applied to a two cell lattice of \textit{M45} for input sequence \((\textit{0},\textit{1},\textit{0},\textit{1})\). Directionality is alternated at each step. Dark nodes and bold edges indicate the current state and transition, respectively. Dashed lines show the composition of inputs to outputs at each step; row outputs are discarded.}
    \label{formulation}
\end{figure}

Four steps in the evolution of a sender/receiver pair of states is shown in \Cref{formulation} for the FSM with identifier \textit{M45}. We adopt the convention of maintaining the position of persisted states by horizontally flipping sender and receiver at each step; this ensures that state changes visually align along columns of the grid and makes transitions easier to follow.\footnote{Thus in this figure, the first and third row have the sender on the left and receiver on the right, while the second and fourth row have the receiver on the left and sender on the right.} It also highlights the spatial symmetry which flipping enforces on the lattice, a mechanism we refer to as ``folding.''

We can now generalize the reflexive composition described above to an arbitrarily-sized lattice. In earlier work, this generalization took the form of repeated application of composition to generate exponentially larger state-space graphs; here our focus will be on a fixed lattice of \(n\) machines, but the update process is identical.

For a lattice of \(n\) machines, the output function is applied across the full set of output functions:

\begin{align}
\label{outputn}
o(t) &= (\varphi_n \circ \varphi_{n-1} \circ \dots \circ \varphi_0)(i(t))
\end{align}

\noindent The transition function generalizes in the same way, by application of state composition and reversal across the full lattice of \(n\) states:

\begin{align}
\label{transitionn}
\begin{split}
q_n(t + 1) &= \delta_0(i(t)) \\
q_{n-1}(t + 1) &= \delta_1(\varphi_0(i(t))) \\
&\dots \\
q_0(t + 1) &= (\delta_n \circ \varphi_{n-1} \circ \dots \circ \varphi_0)(i(t))
\end{split}
\end{align}

\begin{figure}[h]
  \centering
  \includegraphics[width=0.47\textwidth]{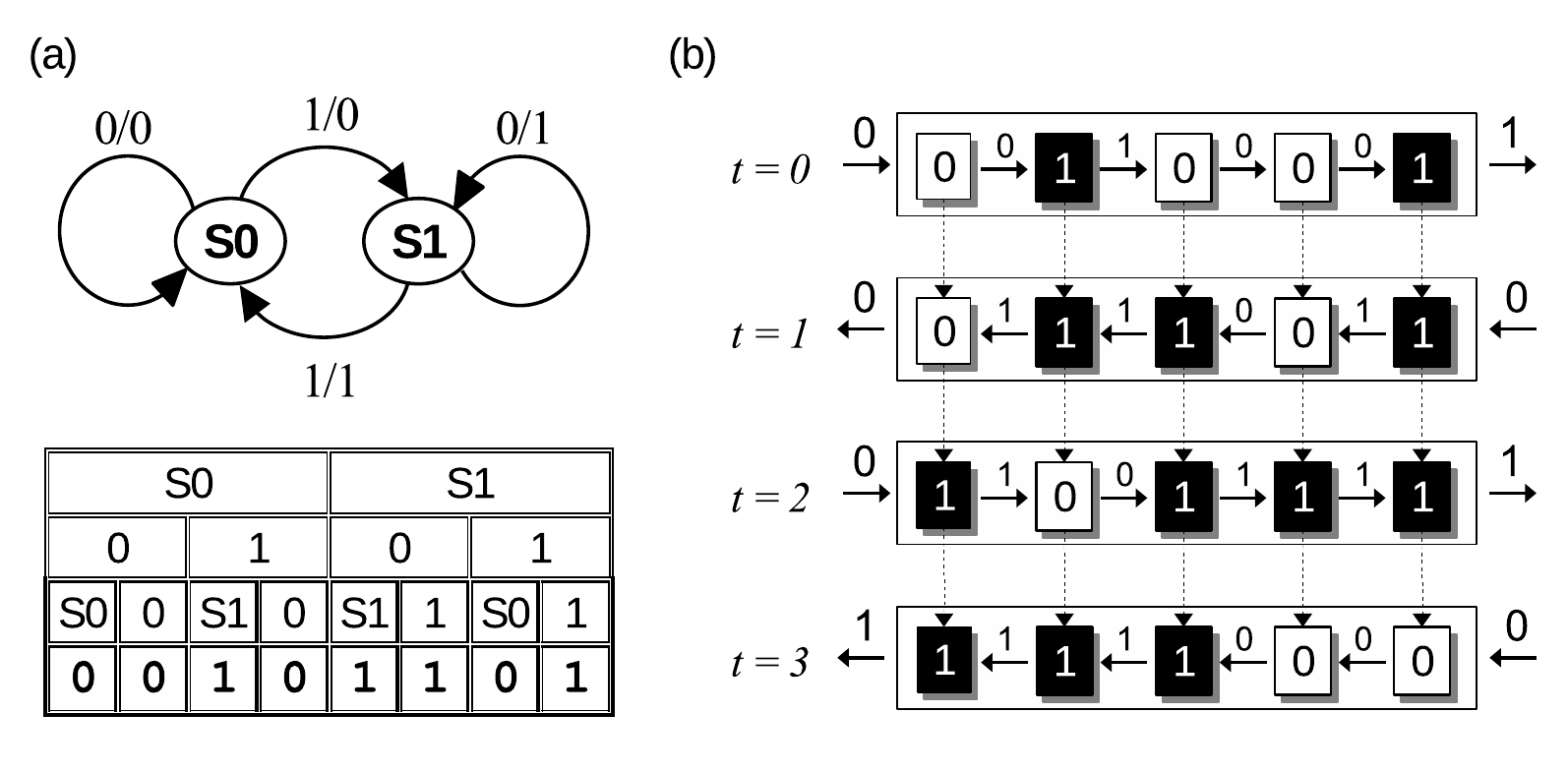}
  \caption{(a) Top: Two-state, two-symbol state machine identified as \textit{M45}. Bottom: table of state machine transitions and derivation of naming (\((\texttt{00101101})_2 = 45\)). Bottom: table of state machine transitions. (b) Four steps of the state composition trajectory of a five-cell \textit{M45} lattice with null-input boundary conditions and initial states \((\textit{0}, \textit{1}, \textit{0}, \textit{0}, \textit{1})\).}
  \label{M45}
\end{figure}

\noindent Equations \eqref{outputn} and \eqref{transitionn} reduce to \eqref{output} and \eqref{transition} when \(n = 2\). Three steps of reflexive composition applied to a 5-state lattice of \textit{M45} machines is shown in \Cref{M45}.

The reader is encouraged to familiarize themselves with the update process shown here; in the next section we omit the details of state transitions at each cell to focus on the dynamics of state changes across much larger lattice sizes.

\section{Results}

\begin{figure*}[h]
  \centering
  \includegraphics[width=\textwidth]{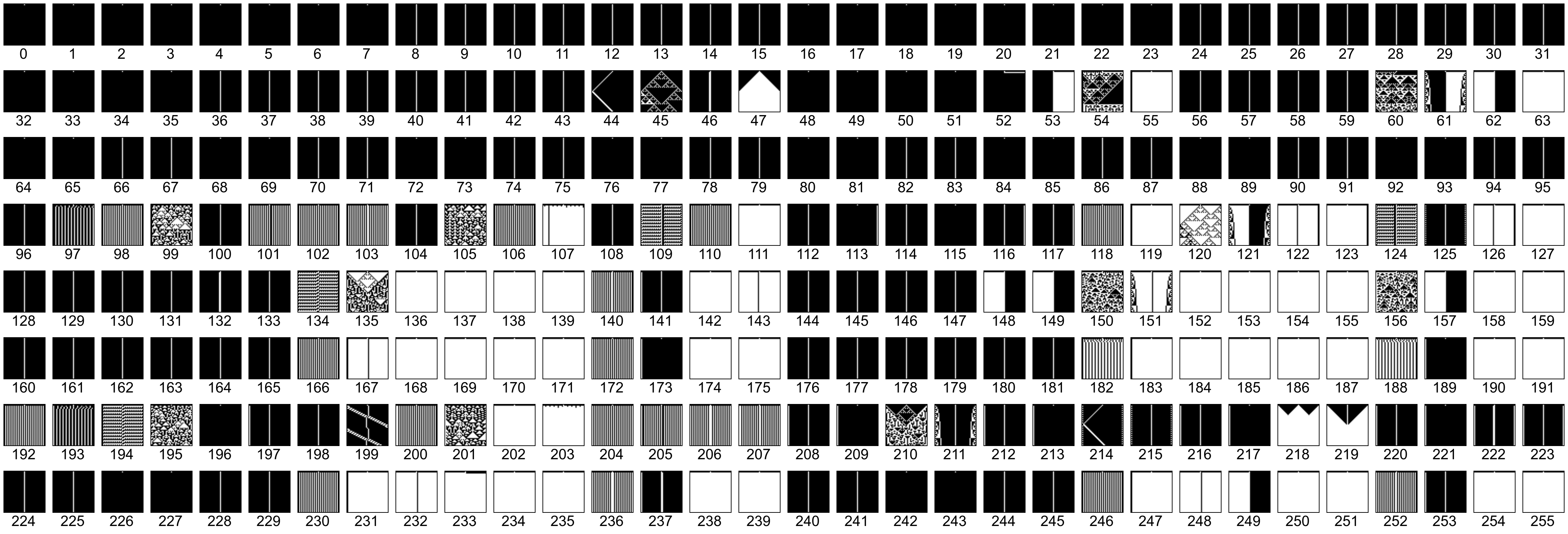}
  \caption{Trajectories of the full collection of 256 2-state, 2-symbol machines from initial centered pixel on 19-cell lattice with input-0 boundary conditions. For visibility only even rows are shown. Black = \textit{S0}, White = \textit{S1}.}
  \label{allmachines}
\end{figure*}

Consider now the full space of elementary FSMs updated using the reflexive composition process described in the last section. \Cref{allmachines} tabulates trajectories for each of these 256 machines starting from a 19-cell lattice with a single centered \textit{S1}-state cell surrounded by \textit{S0}-state cells, with ``\textit{0}-input'' boundary conditions (\(i(t) = 0\) for all \(t\)).

\begin{table}
    \centering
           \footnotesize
    \begin{tabular}{|c|l|l|}
    \hline
\textbf{FSM} & \textbf{Equivalents} & \textbf{Notes} \\
\hline
\textbf{7} & 47, 88, 218 & ECA: R128, R254. \\
\hline
\textbf{44} & 104, 199, 214 & Billiard ball dynamics. \Cref{m44}. \\
\hline
\textbf{45} & 120, 135, 210 & ECA: R90, R165. Figs. \ref{formulation}, \ref{M45} \& 
 \ref{m45r90}. \\
\hline
\textbf{54} & 99, 156, 201 & Reverse R90. Figs. \ref{m54m60}, \ref{m54-dynamics}, \ref{m54m60-dynamics} \& \ref{m54-1200}. \\
\hline
\textbf{60} & 105, 150, 195 & Reverse R90. Figs. \ref{m54m60}, \ref{M54-M60-400} \& \ref{m54m60-dynamics}. \\
\hline
\textbf{61} & 121, 131, 146 & Substitution system. Figs. \ref{M61} \& \ref{m61}. \\
\hline
\end{tabular}
\caption{Selection of elementary FSMs and their equivalent mirrors and complements. There are 76 unique machines in total.}
\label{canonicals}

\end{table}

Although many of the trajectories in \Cref{allmachines} are blank (all \(\textit{S0}\) or \(\textit{S1}\) states), many others exhibit diverse and complex behavior, summarized in \Cref{canonicals}. To help explore this behavior, we begin by partitioning the space of machines in such a way as to clarify its relation to cellular automata.

\subsection{State reporting and message propagation}

We refer to a FSM as \textit{state-reporting} if it satisfies the requirement that it always ``reports its state'', i.e.:

\begin{equation}
\label{statereporting}
\forall i\in\mathscr{A}: \varphi_q(i) = O_q
\end{equation}

\noindent for some constant output \(O_q \in \mathscr{A}\) that is independent of the input \(i\) but unique for different states \(q \in Q\).\footnote{In contrast to its output \(o(t)\), the state \(q(t)\) of such a machine at step \(t\) can be altered by its input at that step. Propagation of messages across the lattice must thus occur through state change in this type of machine.}

An example of a state-reporting FSM is \textit{M45}, shown in \Cref{M45}: regardless of its input, for this machine \(\varphi_{S0}\) returns \(\textit{0}\) and \(\varphi_{S1}\) returns \(\textit{1}\). \textit{M45} thus always conveys its previous state (before transition) to the next cell in the lattice. It is not hard to see that the property of a FSM being state-reporting limits its action at a distance to only directly neighboring cells in the lattice. Consequently, this property is a requirement for equivalence between FSMs under reflexive composition and synchronous cellular automata rules. We show an example of this equivalence later in this section.

The opposite of a state-reporting machine is one which, in at least one case, reports the content of its input in its output. We refer to such machines as \textit{message-propagating} and define them as the negation of \eqref{statereporting}:

\[
\exists q\in Q, i_1, i_2\in\mathscr{A}: \varphi_q(i_1) \neq \varphi_q(i_2)
\]

\noindent In other words, a \textit{message-propagating} FSM is one for which a different input can result in a different output for the same state \(q\). FSMs like this can exert action at a distance and are in general not relatable to synchronous cellular automata. \textit{M61} from \Cref{M61} is an example of a message-propagating FSM since \(\varphi_{S0}(0) \neq \varphi_{S0}(1)\) for this machine.

\begin{figure}
  \centering
  \includegraphics[width=0.45\textwidth]{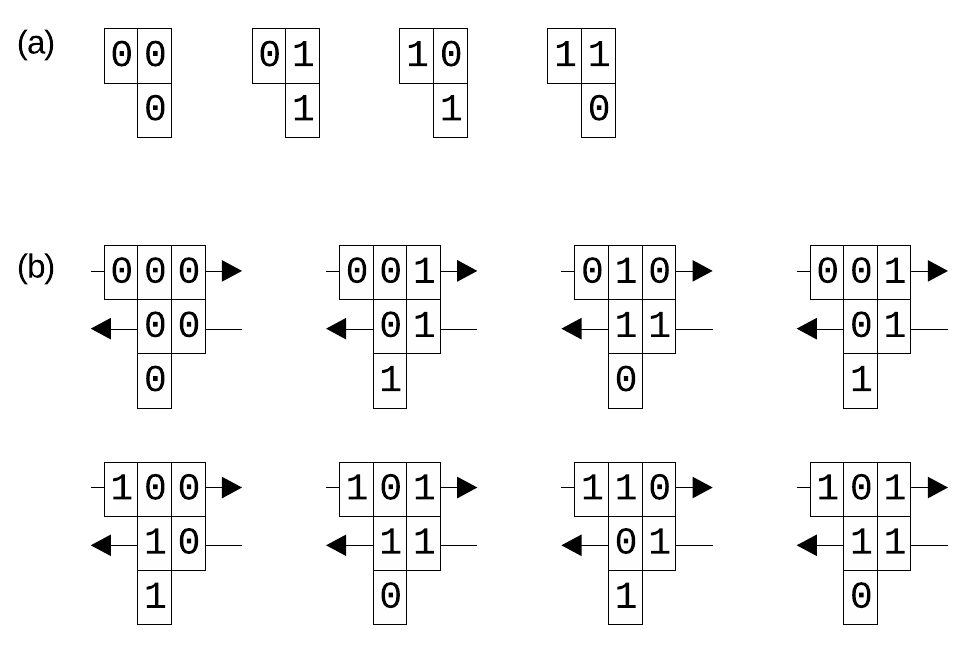}
  \caption{Equivalence between alternate steps of \textit{M45} and each step of CA Rule 90. (a) A simplified view of \textit{M45} purely as a 1-neighbor CA with directionality flipped at each step. (b) Mapping from \textit{M45} transitions to CA Rule 90.}
  \label{m45r90}
\end{figure}

\subsection{Equivalence to cellular automata and M45}

It is easy to show that two steps of a state-reporting FSM under reflexive composition are equivalent to one step of a synchronous ECA. In \Cref{m45r90}, we show that \textit{M45}, seen earlier in \Cref{M45}, is equivalent to CA Rule 90. Intuitively, this equivalence makes sense given that \textit{M45} is the minimal representation of an adding machine modulo 2, and reflexive composition applies this addition twice, once in each direction; this corresponds to the XOR operation of CA Rule 90.

\begin{figure}
    \centering
    \includegraphics[width=0.32\textwidth]{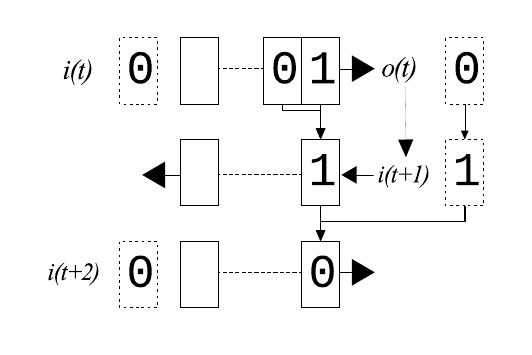}
    \caption{Inputs for \textit{M45} that reproduce null boundaries of R90. Even rows take constant \textit{0} inputs. Odd rows take the output of the previous (even) step as input, corresponding to the value that the ``virtual boundary state'' (dotted boxes) was updated to from its even-row value of \(\textit{0}\).}
    \label{boundaries}
\end{figure}

Boundary conditions in this equivalence relation require special attention, however. We can ensure that the boundaries on alternating steps act like a constant \texttt{0}-valued CA cell by extending the lattice by two ``virtual states'', which act like \textit{S0} states on those steps (but not necessarily on other steps). In \Cref{boundaries}, we illustrate this technique of enforcing CA boundaries on even steps.

The technique requires a different calculation for even and odd steps. Since the left-hand virtual boundary is \(\textit{S0}\), the input to the lattice at step \(t\) is \(\textit{0}\) (since \textit{M45} reports its state). The right-hand virtual boundary cell is updated according to \textit{M45}'s transition function applied to the lattice output \(o(t)\): if \(o(t)\) is \(\textit{0}\), it remains \(\textit{S0}\), if it is \(\textit{1}\), it transitions to \(\textit{S1}\). Since \textit{M45} reports its state, the virtual boundary cell thus sends \(\textit{0}\) as the right-hand input to the next step \(i(t+1)\)  if it remained in state \(\textit{S0}\), otherwise it sends \(\textit{1}\).

This can be condensed to simply:

\begin{align*}
i(t_{even}) &= \textit{0}\\
i(t_{odd}) &= o(t-1)
\end{align*}

\noindent This is equivalent to taking the sum modulo 2 of the previous row (with \textit{S0} and \textit{S1} mapped to 0 and 1, respectively). Applying these boundary conditions to a lattice of \textit{M45} machines with reflexive composition results in a trajectory identical to ECA Rule 90 for even steps. For odd steps, it will diverge from R90 in cases where there are \texttt{1} cells on the boundary of even steps.

\subsection{Message-propagating FSMs}

We have focused so far on state-reporting FSMs, which as shown for \textit{M45}, are equivalent on alternating steps to ECA rules. Message-propagating FSMs, in contrast, can act at a distance across the lattice and are thus not reproducible by synchronous cellular automata. In this section we investigate two interesting cases in this class of FSMs.

\subsubsection{M44: Billiard ball dynamics}

\begin{figure}[h]
    \centering
    \begin{minipage}{0.276\textwidth}
    \centering
    \includegraphics[width=0.8\textwidth]{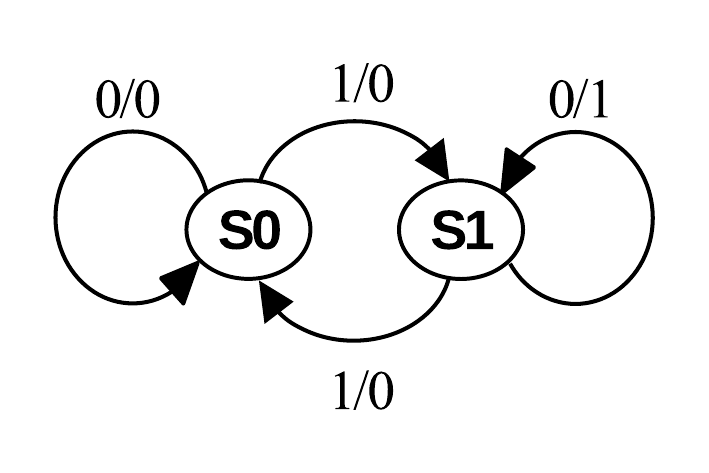}
    \qquad
    \includegraphics[width=\textwidth]{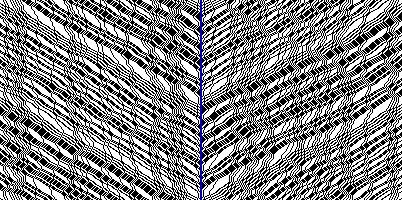}
    \end{minipage}
    \qquad
    \begin{minipage}{0.142\textwidth}
    \centering \includegraphics[width=\textwidth]{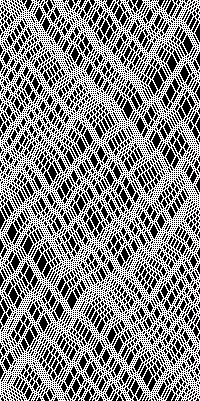}
    \end{minipage}

    \caption{Top-left: \textit{M44} state transition diagram. Right: Four hundred steps of \textit{M44} with initial random 200-state configuration and null input boundaries. Bottom-left: same steps, split into even rows (left) and odd rows (right).}
    \label{m44}
\end{figure}

\Cref{m44} plots the trajectory of \textit{M44} over four hundred steps from an initial random lattice of cells. The dynamics of this rule are reminiscent of invertible cellular automata \citep{Toffoli1990}, which commonly have billiard-ball like dynamics. Indeed, \textit{M44} is reversible and can be inverted simply by flipping the lattice of states at any step in the trajectory.

\subsubsection{M61: Substitution system}

For given initial conditions and boundary inputs, \textit{M61}, shown earlier in \Cref{M61}, produces the patterns of a substitution system along its boundaries \citep{Wolfram2002}.

\begin{figure}[h]
    \centering
    \includegraphics[width=0.47\textwidth]{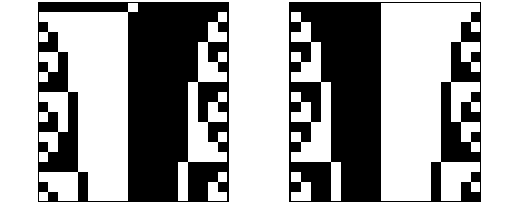}
    \caption{60 steps of \textit{M61}
    with initial configuration of a single centered \textit{1} pixel null-input boundaries, split into even (left) and odd (right) rows.}
    \label{m61}
\end{figure}

\subsection{M54 and M60: Reversal of CA Rule 90}

\begin{figure}
    \centering
    \subfloat[\textit{M54}]{\includegraphics[width=0.193\textwidth]{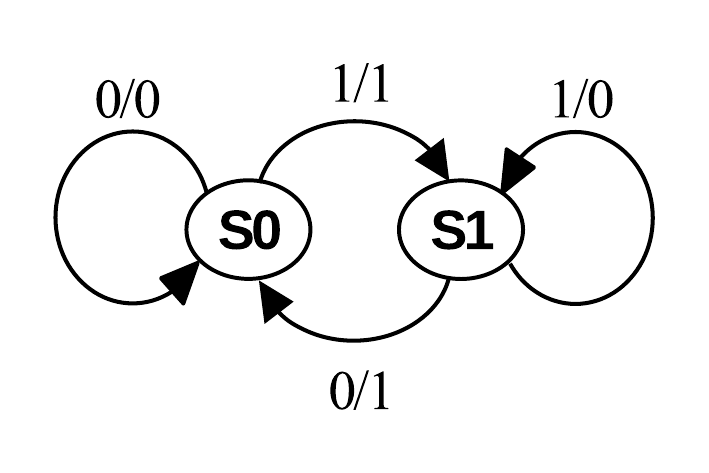}}
    \subfloat[\textit{M60}]{\includegraphics[width=0.193\textwidth]{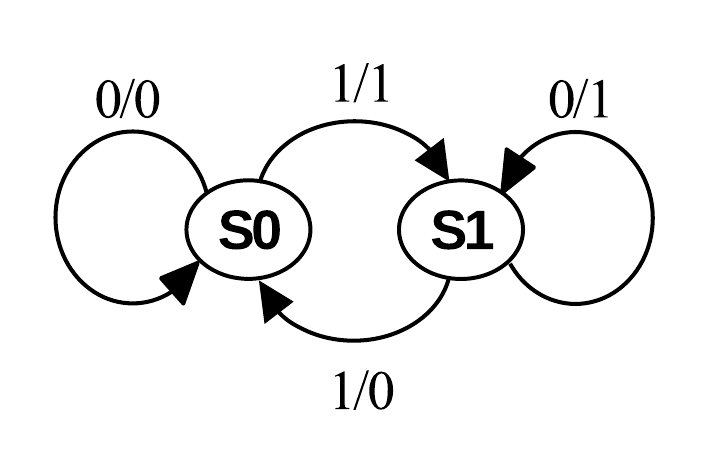}}
    \caption{Transition diagrams of \textit{M54} and \textit{M60}. FSMs differ only in the inversion of their destinations from \textit{S1}.}
    \label{m54m60}
\end{figure}

\begin{figure}[h]
\centering
\subfloat{\includegraphics[width=0.47\textwidth]{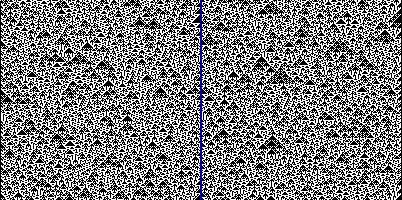}}
\qquad
\subfloat{\includegraphics[width=0.47\textwidth]{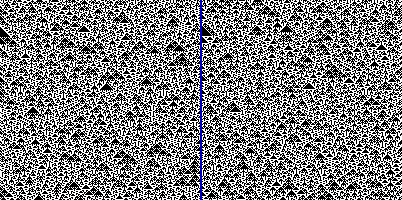}}
\caption{400 steps of \textit{M54} (top) and \textit{M60} (bottom) with the same random initial conditions, split into even (left) and odd (right) steps.}
 \label{M54-M60-400}
\end{figure}

Reflexive composition produces its most intriguing result in the dynamics of message-propagating machines \textit{M54} and \textit{M60}, shown in \Cref{m54m60}. \Cref{M54-M60-400} charts trajectories for (identical) random initial conditions and null-input boundary conditions. The striking complexity of these trajectories stands out among FSMs explored in this study. While reminiscent of ECA, the simultaneous change of state across disparate sites of the lattice is impossible in synchronous CA.

\begin{figure}
    \centering
    \includegraphics[width=0.47\textwidth]{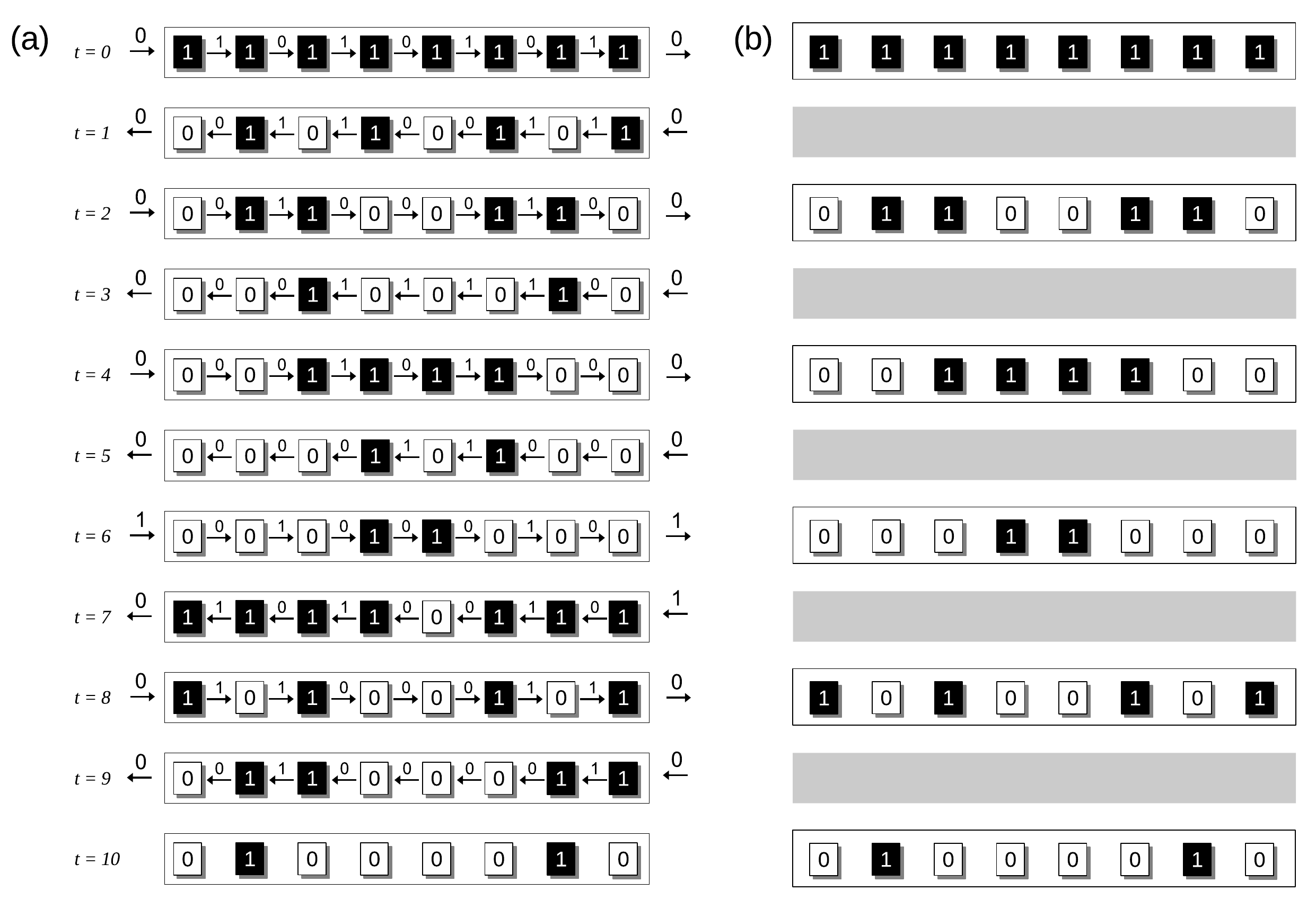}
    \caption{Evolution of \textit{M54} applied to an 8-cell lattice. (a) Even and odd steps. (b) Even steps only, highlighting isomorphism to R90 in reverse. Up to \(t = 5\) inputs are all \textit{0}, but the null constraints of eqs. \eqref{reverseboundary_even} and \eqref{reverseboundary_odd} demand that the input of \(t=6\) and \(t=7\) are both \textit{1} to ensure that \(o(6) = i(7)\) and \(o(7) = 0\).}
    \label{m54-dynamics}
\end{figure}

Closer inspection of the state-change dynamics of \textit{M54} reveals that this machine is in fact the reverse of \textit{M45} described earlier, and as such that it reproduces the reverse of CA Rule 90 (see \Cref{m54-dynamics}.) It is known that R90 is not reversible on a periodic finite lattice if the number of sites in the lattice with value \texttt{1} is odd \cite[p.~227]{Martin1984}. While here we consider null boundary conditions, the same constraint applies to cases where periodicity does not affect results. To avoid irreversible configurations, we thus consider first a symmetric arrangement of initial states in which non-quiescent patterns do not reach the (null) boundaries. As shown in \Cref{m54-dynamics} for an initial condition of two centered \texttt{1} cells, under these conditions \textit{M54} indeed produces the reverse-time trajectory of Rule 90. (Later steps require special treatment of boundaries, see below.)

\begin{figure}
    \centering
    \includegraphics[width=0.47\textwidth]{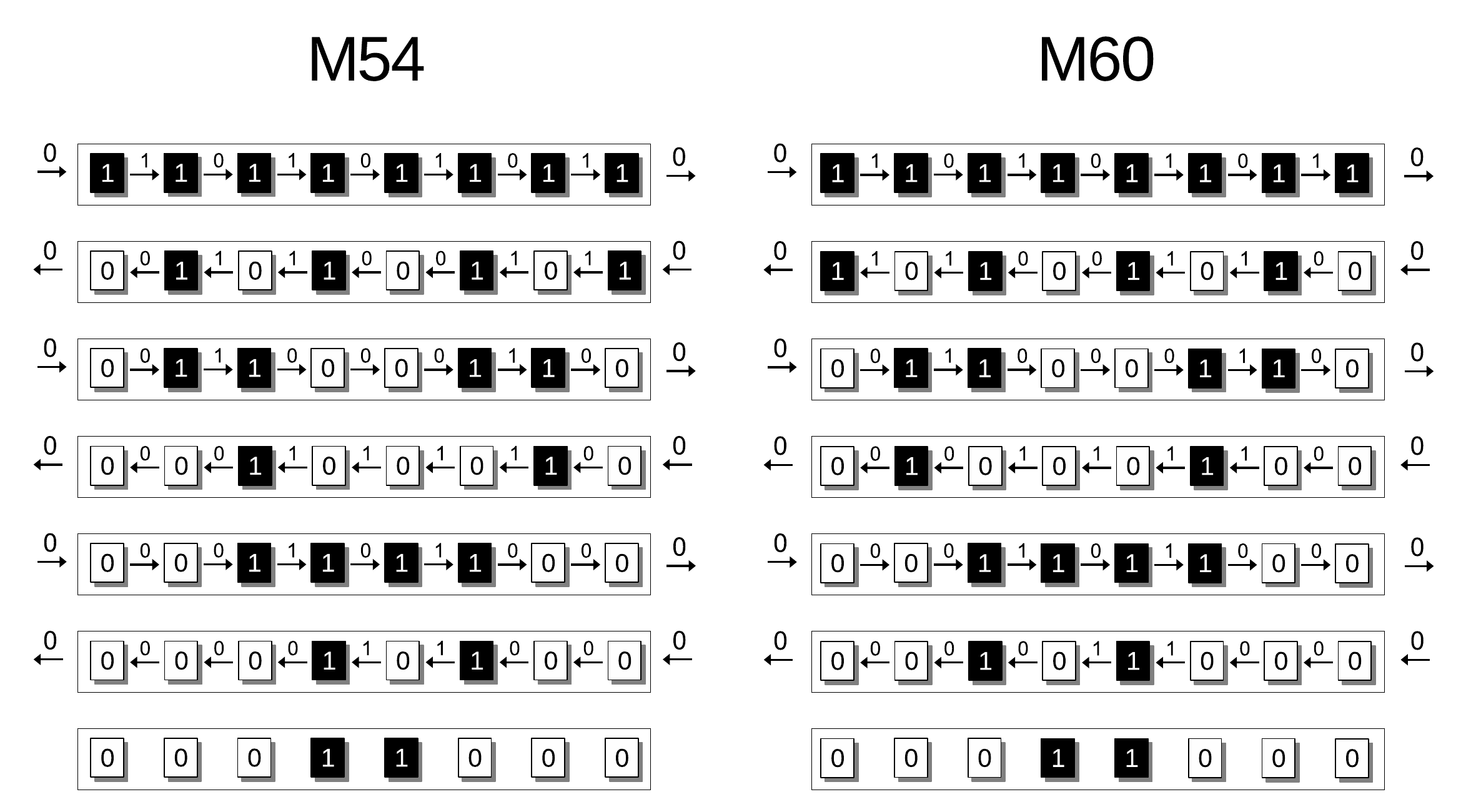}
    \caption{Comparison of \textit{M54} and \textit{M60} for the same initial configuration as \Cref{m54-dynamics}. Even steps are identical; odd steps of \textit{M60} are one cell to the left of \textit{M54}.}
    \label{m54m60-dynamics}
\end{figure}

In \Cref{m54m60-dynamics} we compare the dynamics of \textit{M54} and \textit{M60} with the same initial states and boundary conditions. The lattice configurations of these two machines are identical on even steps and shifted by one cell on odd steps. (We focus below on \textit{M54}, but \textit{M60} produces equivalent dynamics.)

\begin{figure}
    \centering
    \includegraphics[width=0.33\textwidth]{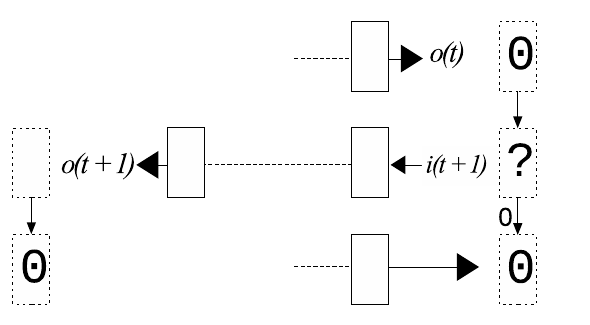}
    \caption{Inputs for \textit{M54} that reproduce null boundaries of R90 in reverse. Output of the even step \(t\), \(o(t)\), transitions the right-hand ``virtual boundary cell'' such that \(i(t+1)\) must be the same value to transition it back to \textit{S0} at the next even step \(t+2\). The odd step output \(o(t+1)\) must be zero to transition the left-hand state to \textit{S0}.}
    \label{m54-boundaries}
\end{figure}

Reproducing reverse dynamics of R90 for the more general case requires correct treatment of boundary conditions. This is shown in \Cref{m54-boundaries}, again using ``virtual'' states to represent null boundaries. We use the fact that \textit{M54} translates its input into the destination state (\textit{0} transitions the state to \textit{S0}, \textit{1} transitions it to \textit{S1}). Thus in order to satisfy the condition that even rows must have a fixed \textit{S0} state starting the row, the output of the previous row \(o(t-1)\) must equal \textit{0}.

Similarly, again using the fact that \textit{S0} translates its input into its destination state, a \textit{0} output for \(o(t)\) will transition the right-hand boundary state to \textit{S0}, and a \textit{1} output to \textit{S1}. The requirement that this transition back to \textit{S0} in the next step constrains \(i(t+1)\) to be \textit{0} if \(o(t)\) was \textit{0} and \textit{1} if \(o(t)\) was \textit{1}.

We can summarize these boundary conditions as:

\begin{align}
\label{reverseboundary_even}
i(t_{odd}) &= o(t - 1)\\
\label{reverseboundary_odd}
o(t_{odd}) &= 0
\end{align}

\begin{figure}
    \centering
    \includegraphics[width=0.47\textwidth]{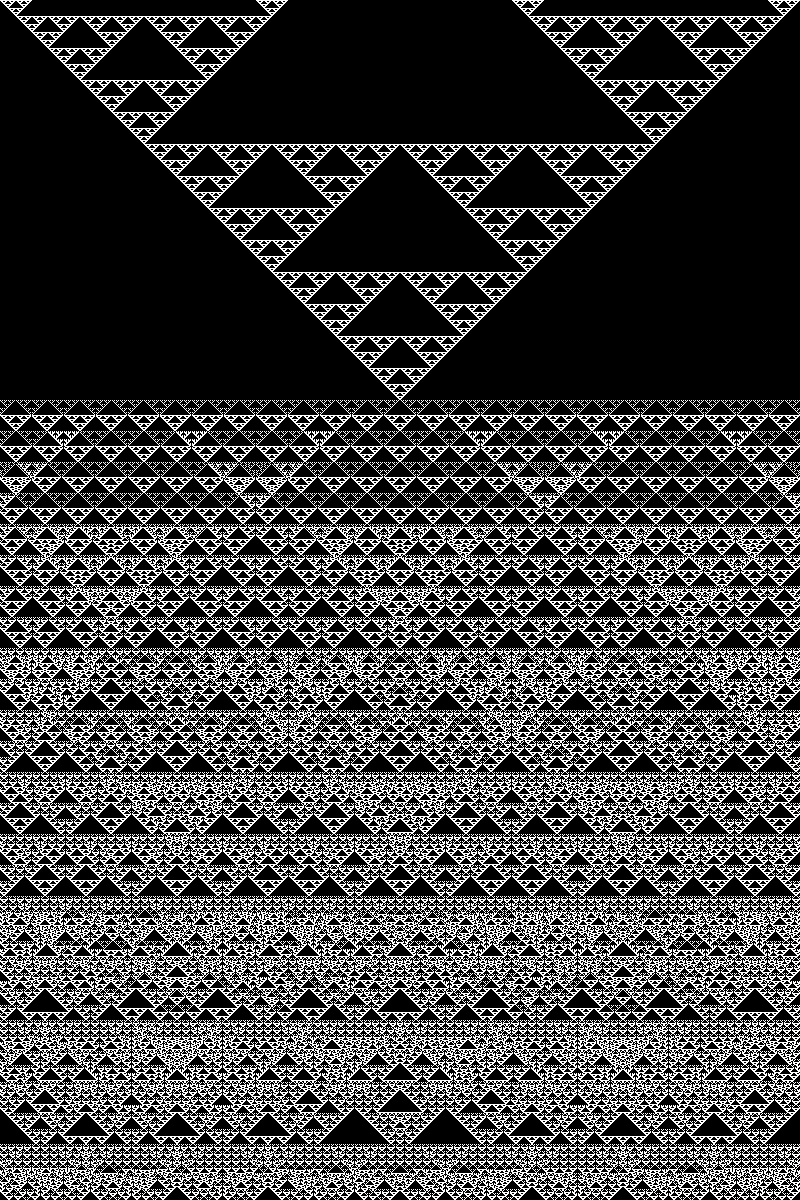}
    \caption{Even steps of 2400-step \textit{M54} trajectory on 800-state lattice initiated with an intermediate R90 configuration. Inputs reproduce forward-direction R90 null boundaries.}
    \label{m54-1200}
\end{figure}

\noindent These constraints are applied using an algorithm in which \(i(t) = \textit{0}\) is first attempted, and if unsuccessful \(i(t) = \textit{1}\) is used instead. This technique is used to generate boundary inputs in \Cref{m54-dynamics}. A larger grid applying these boundary constraints produces the trajectory in \Cref{m54-1200}.

\begin{figure}
    \centering
    \subfloat[\textit{200 cell lattice}]{\includegraphics[width=0.166\textwidth]{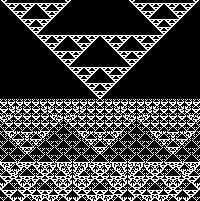}}
    \qquad
    \subfloat[\textit{300 cell lattice}]{\includegraphics[width=0.25\textwidth]{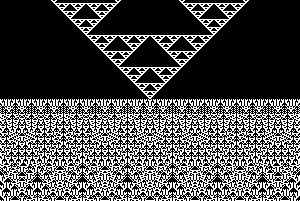}}
    \caption{Even steps of 200-step \textit{M54} trajectory on two lattices of different sizes with the same initial conditions.}
    \label{latticesize}
\end{figure}

While the initial portion of this trajectory is well known, the transition to preimages of the original Rule 90 configuration reveals a striking wealth of complexity not reported, to our knowledge, in any previous studies. The sudden transitioning of entire rows of null cells at once, also exhibited at a smaller scale in \Cref{m54-dynamics}, is made possible by the fact that \textit{M54} is message-propagating. Unlike the forward direction state-reporting machine \textit{M45} (and Rule 90), this reverse direction can propagate messages across multiple cells in a single step; this accounts for coordinated state changes across arbitrary distances in the lattice. The specific arrangement of cells in steps following this transition is moreover strongly affected by lattice size, with even small changes producing entirely different pattern dynamics. \Cref{latticesize} highlights this effect on two lattices with different sizes but identical initial configurations, each producing drastically different state-space trajectories.

\begin{figure}
    \centering
    \includegraphics[width=0.47\textwidth]{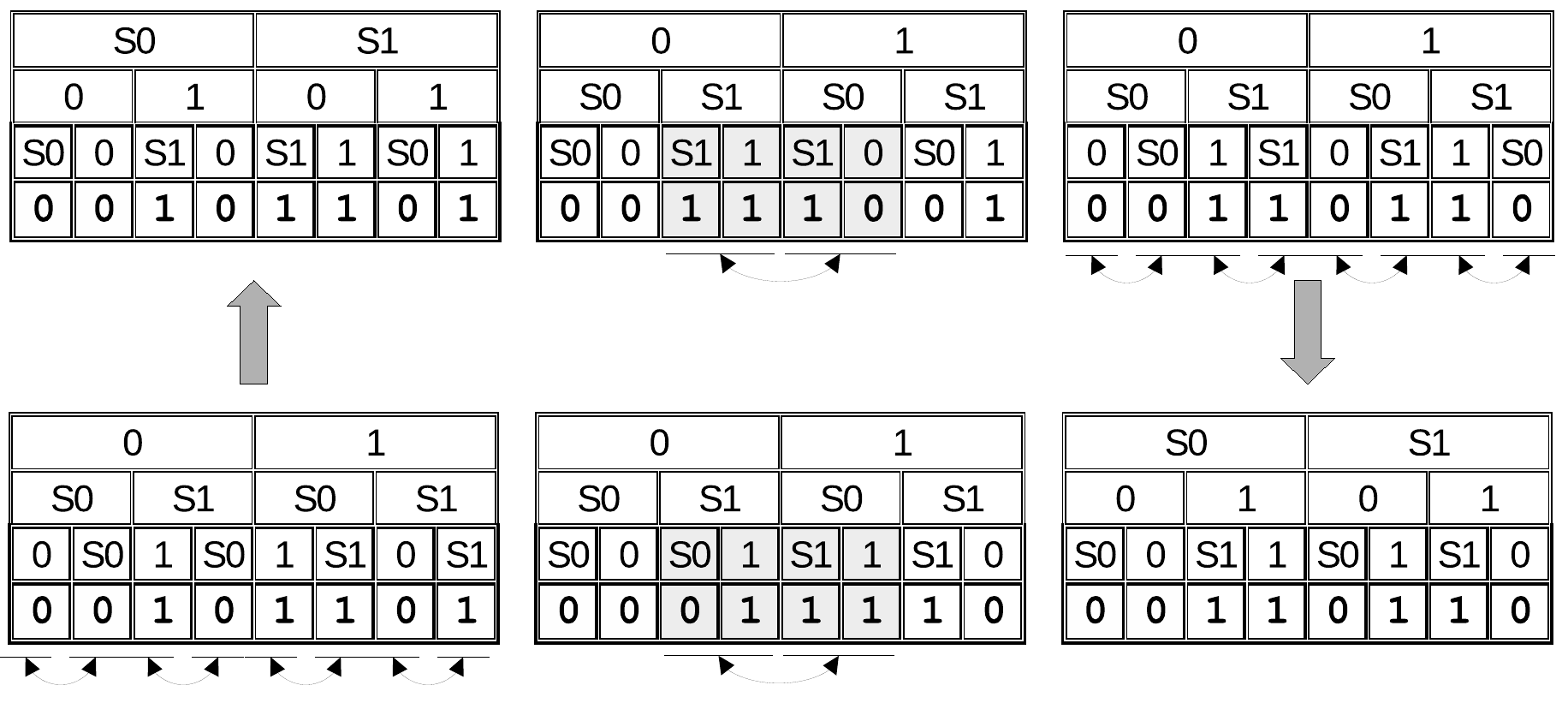}
    \caption{Transposition of state and message transforming \textit{M45} (top left) into \textit{M54} (bottom right), with \(\textit{S0} \leftrightarrow  \textit{0}\), \(\textit{S1} \leftrightarrow \textit{1}\). Moving clockwise, the first step swaps state/message in source and input, the second in destination and output.}
    \label{inverse}
\end{figure}

In contrast to the many techniques for counting and calculating preimages of CA configurations \citep{Jen1989, Voorhees1993, Jeras2007, Stepney2010}, the reverse trajectory of Rule 90 in \Cref{m54-1200} is generated from the selfsame process that produces its inverse. This itself is an unexpected result that hints at a deeper significance. Yet there is a further relationship connecting the forward and reverse direction, arising from reflexive composition's symmetric treatment of \textit{state} and \textit{message}: as shown in \Cref{inverse}, ``transposing'' \textit{M45}  along the state-message axis produces \textit{M54}, and vice versa.\footnote{While not shown here, \textit{M60} transposes into itself and not \textit{M45}. Other machines, notably the billiard ball machine \textit{M44}, do not exhibit this property of state/message transposal resulting in time reversal.} This transposition corresponds to treating messages (inputs and outputs) as states, and states (source and destination nodes) as messages.

What this implies is that the direction of time, for these machines, is fundamentally a product of which elements in the system one chooses to treat as operator and operand. The fact that the two most interesting machines in the set of elementary FSMs exhibit this symmetry hints at the importance of reflexivity in the formulation, and in the unusual dynamics it makes possible. Further investigation of this subtle property and its relationship with complexity is merited.

\section{Conclusions}

We have shown that the process referred to as \textit{reflexive composition} generates a surprising diversity of dynamic behavior from simple origins. In particular, the FSMs identified as \textit{M54} and \textit{M60} reverse the state transition trajectory of CA Rule 90 for configurations that have preimages. Unlike techniques developed specifically to uncover such preimages, here they arise naturally out of the formulation, and indeed in the case of \textit{M45} and \textit{M54} are in fact related to each other through a transposition of state and message.

While Rule 90 is among the most famous and well-studied ECA, state-space trajectories explored in earlier research focus either on familiar patterns and their superpositions or on random initial conditions. In contrast, the reversal of Rule 90 reveals qualitatively distinct patterns containing elements of nested fractal structure of a kind not seen before. We consider this a sign that the underlying formulation presented here, and the origins from which it was developed, manifest deeper truths about the symmetry of information processors and carriers. Future work will expand the scope of exploration to more complex machines and their interactions, with the aim of uncovering these truths and the secrets they hold.

\footnotesize
\bibliographystyle{apalike}
\bibliography{submission}

\end{document}